Extra-auditory Effects of Noise in Laboratory Animals: Focusing on the Relationship Between Noise and Sleep

Arnaud Rabat[1]

[1]Research Department, Institute of Naval Medicine for the French Health Service of Armed Forces, Army Teaching Hospital, Saint Anne, France.

Email: a.rabat@imnssa.net

**Noise has both auditory and extra-auditory effects. Some of the most deleterious extra-auditory effects of noise are those leading to sleep disturbances. These disturbances seem to be related to both endogenous (physical parameters) and exogenous (sex, age) factors of noise. Despite correlative relations between noise level and awakenings, the scientific community has not reached consensus regarding a specific action of these factors on the different sleep stages. In animal research, 2 complementary main fields of research exist. One is focused on the positive modulation of sleep by tone stimulation. The other concerns noise-related sleep disturbances. The few studies that have investigated noise-related sleep disturbances suggest the following conclusions. First, sleep disturbances are greater upon exposure to environmental noise, whose frequency spectrum is characterized by high and ultrasonic sounds, than white noise. Second, unpredictability and pattern of noise events are responsible for extractions from both SWS and PS. Third, chronic exposure to noise permanently reduces and fragments sleep. Finally in chronic noise exposure, an inter-individual variability in SWS deficits develop and are correlated to a psychobiological profile related to an incapability to face stressful situations. Based on results from other research, acute noise-related sleep perturbations could result from an imbalance in the sleep–wake cycle in favor of**

**arousing ascending systems. Chronic noise-related sleep disturbances may arise due to imbalance and malfunctioning of the hypothalamo-pituitary-adrenal axis and may contribute to the development of pathology.**

Abbreviations: PS, paradoxical sleep; SWS, slow-wave sleep

Despite great efforts to reduce noise, it remains one of the main environmental problems of modern society.[48,71] This problem emerged during the end of the 20th century in parallel with the huge development of human activities (airway, railway, and road traffic). According to the World Health Organization, in contrast to many other environmental problems, noise pollution continues to grow, accompanied by an increasing number of complaints from affected individuals.[13] Noise is not simply a local problem—it remains a global dilemma that disturbs everyone and calls for precautionary actions in any environmental planning situation.[13] After briefly summarizing the effects of noise in humans (especially effects on sleep), I will discuss 1) effects of noise on sleep in animals and 2) the neurobiologic mechanisms that explain how noise disturbs sleep.

**Effects of noise in humans**

In humans, noise is now recognized to cause important health problems.[73,92] Noise effects are divided into auditory and extra-auditory changes.[93] Although the auditory effects of noise have long been studied,[15,42] they are not responsible for most of the effects of noise on organisms.[17,92] Noise-associated problems mainly are associated with extra-auditory perturbations, which can appear after to exposure to noise as low as 50 dB[48]. These problems include physiologic (cardiovascular, endocrine), psychologic (mood, attention, memory), and sleep disturbances and can lead to psychiatric problems.[92,93] Further, sleep is altered by acute and chronic exposure to noise.[35,65,66,70] Because of the known restorative function of sleep,[31,113] noise-induced sleep disturbances are the most deleterious effects of noise,[35,38] and many studies in humans have focused on this problem. Both field and laboratory studies have addressed the relation between environmental noise and sleep disturbances. In human subjects in response to peak noise, K complexes on electroencephalograms are accompanied by increases in heart rate, constriction of peripheral blood vessels, and body movements.[38] This initial and typical reaction is followed by more or less long-lasting encephalomyelographic desynchronization indicative of either lightening of sleep or awakening.[38,66] In field studies, correlative relations have been established between complaints or awakenings and noise levels from aircraft or road traffic.[59] Recently, Spreng[90] has proposed a physiologic model to link indoor maximal levels of noise with the number of tolerable noise events during an 8-h night period.[38] Noise-induced sleep disturbances clearly also are related to other endogenous factors of noise, such as the predictability and frequency content of noise events, and to exogenous factors such as gender, age, information content of noise, and individual and situational factors.[35,37,38,65] Indeed, noise-induced sleep disturbances increase with age[59,65,114] and sensitivity to noise.[20] With the same overnight $L_{Aeq}$[1], sleep is more affected by intermittent noises than continuous ones.[20] In addition, sleep disturbances may be linked to the number of noise events.[67] However, there is no real consensus regarding the contribution of each attribute of noise on different sleep stages.[20] In fact, some studies have shown either slow-wave sleep (SWS)[84] or paradoxical sleep (PS) disturbances[30,34,52,97,103] in humans after continuous exposure to noise, whereas others have demonstrated both SWS and PS perturbations.[24,41,105] Some authors found SWS disturbances in human subjects exposed to intermittent noise,[30,36,102] whereas others demonstrated disruptions in either PS only[33,86] or both PS and SWS.[52,98] What happens in laboratory animals that are exposed to noise? Can we draw the same conclusions as for humans?

---

[1] referred to the equivalent sound level ($L_{Aeq, T}$) which determines the global level of noise over a period of time, T, according to this formula: $L_{Aeq,T} = 10 * Log(\frac{1}{T})\int_{0}^{T} 10^{L(t)/10} * dt$.

**Noise and sleep: studies in laboratory animals**

Traditionally, research using animal models delves deeply into questions that arise during research involving human subjects. In noise research, laboratory animals have been used to determine how noise disturbs body and brain functions,[25] with the majority of works focusing on auditory damage. As in humans, auditory damage in laboratory animals appears upon prolonged or repetitive exposure to intense noise (that is, greater than 85 dB sound pressure level). However few animal studies have addressed noise-related sleep disturbances.

Khazan and Sawyer[53] were the first to demonstrate, in rabbit, disruption of PS upon continuous exposure to white noise (78 dB). Initially, the purpose was to demonstrate a PS rebound effect after PS deprivation. Van Twyver and colleagues[108] demonstrated PS disturbances in rats continuously exposed to intense white noise (that is, 92 dB). These 2 studies revealed a direct and specific effect of continuous exposure to white noise on PS that leaves SWS intact. These results were confirmed in a recent study in which the authors compared the effects on sleep in rats of acute exposure (24 h) to either environmental noise or to different kinds of white noises (continuous, intermittent).[78]

Many other studies have used noise or tones to disturb sleep in laboratory animals, but their purposes were not to determine how the noise in laboratory animal facilities disturbs sleep. For example, Van Twyver and Garrett[107] determined, using a meaningful tone (associated with electric footshock), that arousal thresholds were highest during PS, low during SWS, and lowest during waking; these results were partially comparable to those of humans studies.[16,81] Drucker-Colin and colleagues[29] wanted to determine whether ponto-geniculo-occipital spikes play a role in triggering or maintaining sleep. To this end the researchers repetitively exposed (every 20 s) cats to a high-intensity tone (2 kHz at 90 dB) and noted that the tones had a stimulating effect on these spikes, accompanied by increases in the duration and number of PS epochs. The same researchers later demonstrated, using similar experimental conditions, a direct stimulating effect of tones on the duration of PS epochs,[4] with no habituation effect.[109] The increase in the duration of PS epochs after tone stimulation was shown to be dependent on activation of the medial pontine reticular formation.[3] Ball and colleagues showed a dose-related effect of tone intensity on the ponto-geniculo-occipital waves elicited in cats.[9] These results were confirmed in 2 human studies[64,87] and in a study using young and aged rats.[5]

Other recent studies in rats yielded more detailed conclusions. Amici and colleagues[1,2] found that auditory stimulation with repetitive tones (1 kHz during 20 ms every 20 s with different intensities from 50 to 100 dB) induced a significant increase in the duration of PS epochs. This effect was induced with a lower intensity threshold when tone stimulation was applied during SWS than during PS.[2] In a complementary way, Velluti and Pedemonte studied the interaction between sleep regulation and auditory perception.[110,111] Pedemonte and colleagues demonstrated an increase in both SWS and PS epochs after deafness in guinea pigs[74] and concluded that the auditory system may influence central nervous system structures implicated in sleep regulation and thus may contribute to homeostasis of the sleep–wake cycle.[74]

Regarding interactions between noise and sleep in laboratory animals, one broad area of research addresses the positive modulation of sleep, especially PS, by repeated tone stimulations. Such stimulation increases sleep, particularly PS, with no habituation.[3-5,29] PS increases most when stimuli are applied during SWS epochs.[1,2] Amici and colleagues interpreted these results to indicate differences in sensory processing between the sleeping brain in a SWS state and the sleeping brain in a PS state.[2] This result is not surprising, because awakening thresholds in rats are lower during SWS than during PS.[107] This hypothesis is consistent with humans studies demonstrating that 1) the sleeping brain can process auditory stimuli and detect meaningful events[77,109] and 2) awakenings thresholds are lower during SWS than during PS.[16,81]

Another key area of noise research in laboratory animals concerns noise-related sleep disturbances.[53,78,79,108] Until very recently,[78,79] no studies had addressed this problem. In a pilot study, Rabat and colleagues[78] found that sleep in rats was disturbed to a greater extent under exposure to environmental noise (EN²)—which (once adapted to rat audition) is very close to noise encountered in laboratory animal facilities[104]—than under white noise. Indeed, with continuous white noise, intensity shortens the duration of SWS epochs and indirectly the number of PS ones whereas intermittent noise events extract rats from both SWS and PS epochs and induce a stronger and longer effect than does continuous noise. These findings are in agreement with previous animal studies[53,108] as well as with human studies implicating a role of continuous noise exposure in PS deficits[30,34,52,97,103] or in both SWS and PS perturbations[24,41,105] and of intermittent noise on SWS disturbances[30,36,102] or disruption of either PS[33,86] or both PS and SWS.[52,98]

From these animal studies the following 3 conclusions can be proposed. First, both environmental and white noise disturb sleep,[53,78,108] but environmental noise is more deleterious to sleep than is white noise even if its global noise level ($L_{Aeq}$) is higher than that of environmental noise.[78] Two physical components of environmental noise are responsible for its effect on sleep:[78] the unpredictability of noise events and their pattern, such as frequency content and the ratio between background and peak levels. These 2 physical components do not prevent rats from entering SWS and PS, as does continuous white noise, but awakens them from all sleep stages. Interestingly, Stanchina and colleagues[91] recently demonstrated that arousal decreased substantially in an environment with intensive care units noises combined with continuous moderated white noise (62 dB) compared with this type of environmental noise alone. The authors concluded that the difference in intensities between peak and background noise (known as 'delta dB') must be higher than the arousal threshold for waking to occur. As suggested by Ising and Kruppa[48] and confirmed experimentally,[78,79] most important to tone intensity is the unpredictability of noise peak and the pattern of noise especially with its information content, such as the presence of high and ultrasonic frequencies, often present in many laboratory animal rooms.[63,85]

---

² Corresponding to noises that have major octave band frequencies higher than 1,000 Hz.

The second conclusion regarding sleep and exposure to noise in laboratory animals is that upon chronic exposure to environmental noise, both SWS and PS are altered permanently.[79] Again the unpredictability and pattern of noise events seem to be implicated in these effects.[79] Some human studies have paralleled these findings, showing a stronger and longer deleterious effect of intermittent noise events on sleep compared with those of continuous noise.[20,30,68,101,102]

The third conclusion regarding sleep and exposure to noise is that extra-auditory noise effects do not affect all animals equally. Indeed Rabat and colleagues[79,80] recently showed that under chronic exposure to noise, some rats accumulated a bigger SWS debt with greater fragmentation of the remaining SWS than did other rats. In this work, locomotor reactivity of rats placed in a novel environment, before their exposure to noise, was correlated with both the intensity of SWS debt (a quantitative measure of sleep) and daily decrease of SWS bout duration (a qualitative parameter). Behavioral reactivity to novelty predicted quantitative and qualitative characteristics of SWS disturbances related to environmental noise and reflected the existence of separate psychobiologic profiles (high-reactive versus low-reactive) in rats[27,28] with neurochemical, neuroadaptive, and neuroendocrine characteristics.[44,51,57,75] Interindividual vulnerability to noise-induced sleep alterations may reflect these specific neurochemical and neuroendocrine profiles. Indeed, noise exposure modifies neurochemical transmission in the brain.[6,55,56,99] Furthermore, high-reactive and low-reactive rats showed distinct basal modifications in cholinergic, dopaminergic, and serotoninergic systems in the brain.[44,75,100] Some authors previously had suggested interindividual vulnerability in humans to auditory[26] and extra-auditory damage.[10,11,20,49,114] This last point brings us to the last question addressed in this review: how does noise act on the brain to cause the described sleep disturbances?

**An experimental hypothesis to explain how noise disturbs sleep**

As a sensory stimulation, noise can induce a wide range of responses that allow the organism to gather information from the source of the stimulation and to develop an adapted behavioral response. Thus noise engages peripheral auditory system and, in a general way, the central nervous system (Figure 1).

Lai and colleagues[55,56] demonstrated that when rats were exposed acutely to white noise of low intensity (70 dB), choline recapture in the frontal cortex and hippocampus of rats was increased, but this recapture was decreased with white noise of higher intensity (100 dB). Choline recapture is a direct reflection of acetylcholine (Ach) transmission.[39,40,60].

In addition, pontine structures, such as the reticular formation and pedunculopontine nucleus, receive auditory input[82] and modulate the cholinergic activities of basal ganglia, limbic, thalamic, and hypothalamic structures,[54,76,89] all of which are implicated in sleep regulation[12,43]. Cholinergic neurotransmissions in pontine structures and the basal forebrain thus might contribute to disturbing sleep after acute exposure to noise (Figure 1). For example, a recent study found that after total sleep deprivation for 6 h, mice null for the dopamine β-hydroxylase gene, which encodes an enzyme involved in the production of norepinephrine in the brain, show a significant increase in acoustic arousal threshold compared with that in wild-type mice[46]. The authors hypothesized that norepinephrine-containing neurons of the locus coeruleus, one of the wake promoting regions[88] activated by noise,[18] in the knockout mice would be inefficient and thus contribute to imbalance the sleep-wake cycle in favor of sleep[46]. NE therefore may be another neurotransmitter responsible for noise-related sleep disturbances (Figure 1). Because noise promotes waking by activating many pontine and brainstem arousal systems,[45,89] acute sleep disturbances could be due to a temporary and reversible imbalance of sleep neurotransmissions in favor to wakefulness (Figure 1). In the situation of chronic noise exposure, this hypothesis is necessary but not sufficient to explain long-lasting sleep disturbances.[50,72]

Repetitive or chronic noise exposure induces a neurotropic response with corticosterone secretion.[7,32,47] This hormone exerts inhibitory effects on SWS state,[21-23] and sleep-deprived human subjects have elevated levels of glucorticoids.[58] In addition differing corticosterone responses to stressful situations have been related to distinct psychobiological profiles.[27] In a recent work, feedback regulation of the hypothalamo-pituitary-adrenal axis under basal conditions was quite different between high-locomotor-responder-to novelty and low-locomotor-responder-to novelty rats[51]. Further, Rabat and colleagues[79] showed that interindividual vulnerability of SWS to chronic noise exposure were predicted by the psychobiological profile. Taken together these results suggest that, in situations of chronic noise exposure, malfunctioning of hypothalamo-pituitary-adrenal axis could occur and, when accompanied by disruption of sleep–wake neurotransmissions, could be sufficient to induce chronic sleep disturbances and explain interindividual vulnerability. Figure 1 provides a more detailed explanation of how acute and chronic exposure to noise disrupts sleep. Figure 2 presents the flip–flop switch model which highlights the delicacy balance in the regulation of the sleep–wake cycle.

**Conclusions, recommendations, and perspectives**

Despite the paucity of research, noisy environments, such as those encountered in laboratory animal facilities,[104] potentially contribute to sleep disturbances in animals. Recent studies argue against habituation to such exposure to environmental noise. Two physical components are mainly responsible: the unpredictability of noise events and their spectral composition (especially when they are high-frequency [greater than 4 kHz]). Furthermore, individual laboratory animals may not be affected equally by noise-related sleep disturbances. Offline information processing occurs during both SWS and PS,[12,43,83,95,96] as does a restorative brain process during SWS;[113] therefore noise-induced sleep restriction might lead to interindividual vulnerability in cognitive deficits. Recent data from rats,[80] confirmed by studies in humans,[106] support this hypothesis.

Because SWS is intimately linked to both activity of the hypothalamo-pituitary-adrenal axis[94] and of the immune system,[61,69] animals chronically exposed to noisy laboratory animal facilities may develop pathology.[62] To avoid such health problems, managers and veterinarians should ensure that acoustic environment of laboratory facilities does not contain excessive high-frequency noises. These personnel also should ensure that the absolute difference between peak and background levels (delta dB) does not exceed 30 dB.

Future research should verify the accuracy of described hypotheses regarding noise-induced pathology and identify measures to limit or counteract such consequences.

## Acknowledgments

This review is dedicated to Jean-Jacques Bouyer. Thanks to Martine Kharouby for her technical assistance. Thanks to Jean-Marie Aran, Alain Courtière, Olivier George, Muriel Koehl, Monique Vallée, Willy Mayo, and Michel Le Moal, who contributed in varying degrees to the works cited.[78-80] All of the work described was completed under the research units "Psychopathologie des comportements adaptatifs" (Inserm U259) and "Physiopathologie du comportement" (Inserm U588) at Bordeaux and was supported by grants from Inserm, Université Victor Segalen Bordeaux 2, and Délégation Générale à l'Armement (D.G.A.).

**Figures captions**

**Figure 1.** A simplified theoretical model of how acute and chronic exposure to noise induces sleep disturbances Noise interacts with auditory systems (gray boxes; PAS, peripheral auditory system; IC, inferior colliculus; MG, medial geniculate body; AC, auditory cortex; FC, frontal cortex; TpC, temporal cortex) to produce both conscious detection of the stimulus and auditory effects if noise is too loud or prolonged. Concurrently acute and chronic noise activate, in the thalamus, the medial division of the medial geniculate body (MMG). This nucleus, which connected to the lateral nucleus (LA) of the amygdala (1, green line), the gateway of noise into neuronal systems implicated in wake–sleep regulation (red and blue boxes) and adaptive responses to emotional and stressful stimuli (green boxes).

The first consequence of acute exposure to noise is activation of the central nucleus of amygdale (ACE) of both the locus coeruleus (LC) and of 3 cholinergic nuclei (pedunculopontine [PPT], laterodorsal [LDT], and basal forebrain [BF] nuclei; 2, green lines). These serotoninergic (5HT) and cholinergic (Ach) neuronal systems promote wakefulness to the detriment of sleep (3, red line). This wake-promoting effect is amplified by inhibition of the ventrolateral preoptic nucleus (VLPO) of the posterior hypothalamus by monoaminergic nuclei (LC, DR, and TMN) of the ascending arousal system (4, red line). This inhibition (see Figure 2 for additional details) relieves inhibition of the lateral hypothalamus (LH; 5, blue line) that exerts, through hypocretins (Hcrt) and melanoconcentrating hormone (MCH) neurons, a reinforcing effect on all ascending arousal systems (5, red line). All of these factors contribute to increase wakefulness in animals exposed to noise.

The second consequence of acute exposure to noise is activation of the ventrolateral part of the periaqueductal gray (vlPAG) by ACE (6, green line). This nucleus connected to the lateral (lPAG) and dorsolateral (dlPAG) parts of the PAG, induces many body and brain responses (7, green lines), including physiologic (increases in heart and respiratory rates and blood pressure), behavioral (fight-or-flight response to acute exposure to noise and loss of control in a chronic exposure to noise), and endocrine (secretion of catecholamines, epinephrine, norepinephrine, corticosterone, and testosterone) effects. With ACE, corticosterone activates the paraventricular nucleus (PVN) of the hypothalamus (8, green lines). This nucleus contributes to increasing wakefulness to the detriment of sleep by activating LC (9, green line). In a reciprocal manner, LC activates PVN (9, Red line).

Under chronic exposure to noise, the effects described for acute exposure would be intensified. Decreased amounts and noise-induced fragmentation of sleep could result from chronic activation of ascending arousal systems and would be counteracted in part by circadian and sleep homeostatic input. Prolonged secretion of corticosterone may lead to malfunctioning of the hypothalamo-pituitary-adrenal (HPA) loop, as characterized by amplification of both corticosterone secretion and PVN activation (10, green lines). Consequently, activation of ascending arousal systems would be maintained and amplified in a feed-forward loop (9, green and red lines). Furthermore, chronic physiologic, hormonal, and behavioral changes (especially loss of control, because animals strive to free themselves from noise exposure) would contribute to many detrimental health consequences such as cardiovascular, depression-like, and immunologic effects (11, black lines). Inspired from 8, 14, 19, 47, 72, 88, 112.

**Figure 2.** A simplified description of the flip–flop switch model proposed by Saper and colleagues.[88] (A) In this model, wakefulness emerges after inhibition of the ventrolateral preoptic nucleus (VLPO) by monoaminergic nuclei (locus coeruleus [LC], dorsal raphe [DR], and tuberomammillary nucleus [TMN]). This inhibition relieves inhibition from VLPO to the lateral hypothalamus (LH), which has hypocretins (Hcrt) and melanoconcentrating hormone (MCH) neurons that exert a reinforcing effect on all of the ascending arousal system (pontine and brainstem system). (B) During sleep, firing of VLPO neurons leads to production of γ-aminobutyric acid (GABA) and galanine (Gal); these neurotransmitters inhibit

monoaminergic nuclei (LC, DR, and TMN) to counteract VLPO inhibition. In addition, VLPO neurons inhibit the Hcrt and MCH systems of the LH, thus preventing monoaminergic activation that might interrupt sleep.

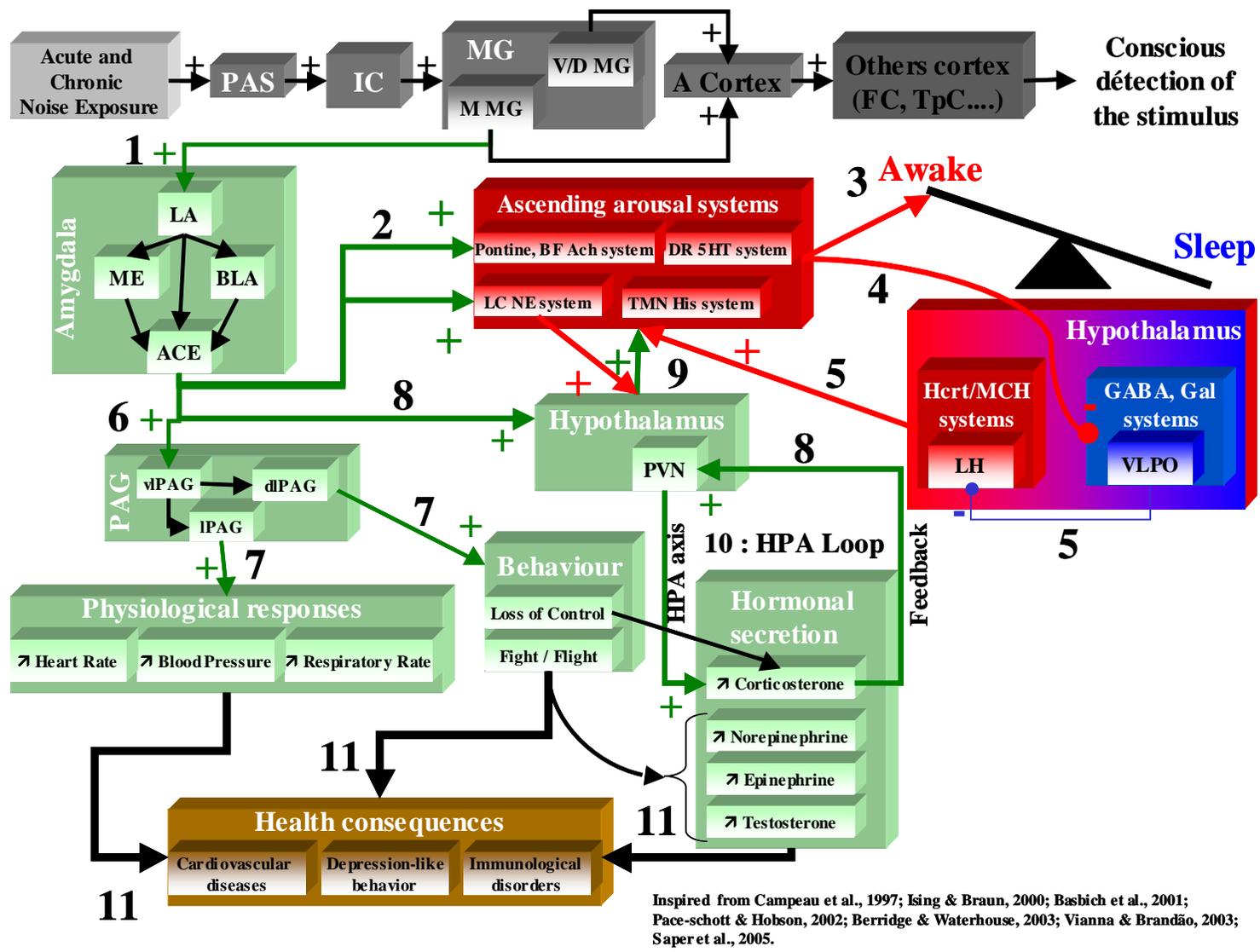

**Figure 1**

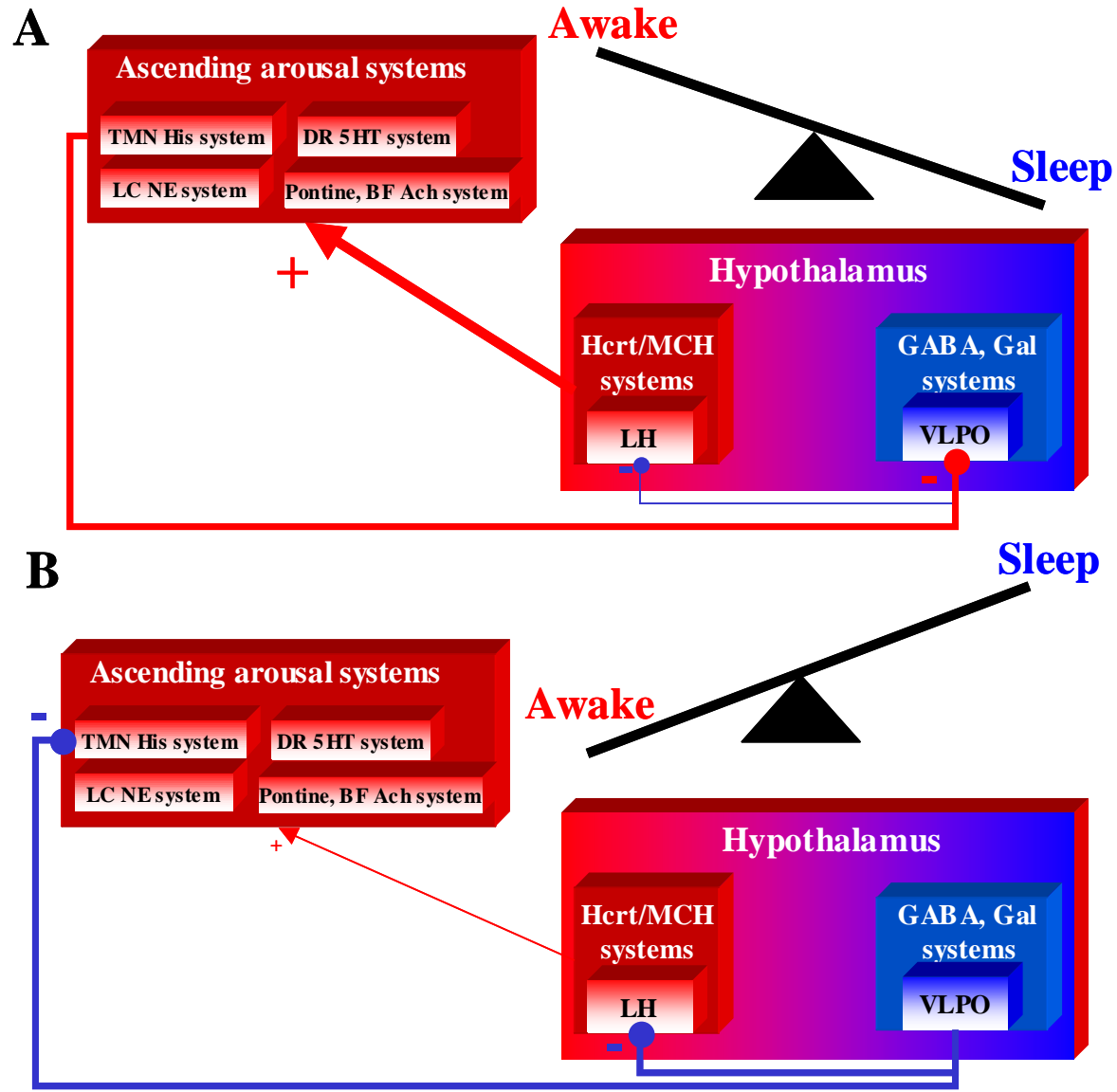

**Figure 2**

Adapted from Saper et al., 2005